\documentclass[aps,prl,reprint,groupedaddress,nofootinbib]{revtex4-1}
\usepackage{graphicx, color, soul}
\usepackage{xcolor}      
\usepackage{siunitx}
\usepackage{amsmath}
\usepackage{braket}
\usepackage{color,soul,url}
\usepackage{float}

\begin{document}
\title{Observation of Quadratic (Charge-2) Weyl Point Splitting\\ in Near-Infrared Photonic Crystals}
\author{Christina J\"org$^{1,2,\dag,*}$, Sachin Vaidya$^{2,\ddag,*}$, Jiho Noh$^{2,3}$, Alexander Cerjan$^{2,4,5}$, Shyam Augustine$^{1}$, Georg von Freymann$^{1,6}$, Mikael C. Rechtsman$^{2}$}
\affiliation{
 $^1$Physics Department and Research Center OPTIMAS, Technische Universität
Kaiserslautern, 67663 Kaiserslautern, Germany\\
 $^2$Department of Physics, The Pennsylvania State University, University Park, PA 16802, USA\\
 $^3$Department of Mechanical Science and Engineering, University of Illinois at Urbana-Champaign, Urbana, Illinois 61801, USA\\
 $^4$Sandia National Laboratories, Albuquerque, New Mexico 87123,  USA\\
 $^5$Center for Integrated Nanotechnologies, Sandia National Laboratories, Albuquerque 87123, New Mexico, USA\\
 $^6$Fraunhofer Institute for Industrial Mathematics ITWM, 67663
Kaiserslautern, Germany\\
 $^*$These authors contributed equally to this work.
}

\def\thefootnote{}\footnotetext{$^\dag$ christinaijoerg@gmail.com, $^\ddag$ sxv221@psu.edu}

\date{\today}

\begin{abstract}
Weyl points are point degeneracies that occur in momentum space of three-dimensional periodic materials and are associated with a quantized topological charge. We experimentally observe the splitting of a quadratic (charge-2) Weyl point into two linear (charge-1) Weyl points in a 3D micro-printed photonic crystal via Fourier-transform infrared spectroscopy. Using a theoretical analysis rooted in symmetry arguments, we show that this splitting occurs along high-symmetry directions in the Brillouin zone. This micro-scale observation and control of Weyl points is important for realizing robust topological devices in the near-infrared.
\end{abstract}
\maketitle

\section{Introduction}
Weyl materials have drawn significant interest over the last decade for their ability to support Weyl points (WPs), which are topologically protected degeneracies in three-dimensional periodic systems. They have been examined in a wide range of physical systems such as conventional solids~\cite{xu_discovery_2015,PhysRevX.5.031013,lv_observation_2015,PhysRevLett.117.146401,RevModPhys.90.015001,MENG2020523,doi:10.1146/annurev-conmatphys-031016-025458}, photonic systems~\cite{lu_weyl_2013,Lu622,Yang1013,noh_experimental_2017,PhysRevB.95.125136,https://doi.org/10.1002/lpor.201700271,JPSJ.87.123401,cerjan_experimental_2019,Sachin,Wang271,devescovi2021cubic}, acoustic systems \cite{PhysRevLett.117.224301,ma2019topological,deng_acoustic_2020,luo_observation_2021}, cold atoms~\cite{Mai_2017,PhysRevLett.114.225301}, electric circuits~\cite{Li_2020} and coupled resonator arrays~\cite{lin_photonic_2016,PhysRevA.96.013811}. In photonics, WPs are sought for a range of applications such as designing large-volume single-mode lasers \cite{Chua:14} and mediating long-range interactions between quantum emitters  \cite{PhysRevLett.125.163602, PhysRevLett.123.173901}, and as such they have been realized in a variety of optical and photonic systems~\cite{lu_weyl_2013,Lu622,Yang1013,noh_experimental_2017,PhysRevB.95.125136,https://doi.org/10.1002/lpor.201700271,JPSJ.87.123401,cerjan_experimental_2019,Sachin,Wang271,devescovi2021cubic} spanning orders of magnitude in wavelength. WPs are quantized sources and sinks of Berry curvature in momentum space, analogous to magnetic monopoles. Therefore, they can be topologically characterized by the closed surface integral of the Berry curvature; this ``charge" is always an integer and is identical to the first Chern number. On the 2D surfaces of a Weyl material, this topological charge manifests through the appearance of surface-localized conduction channels, so-called Fermi-arc surface states, that connect two WPs with opposite charges. Since the Berry curvature vanishes identically in the Brillouin zone for structures with both spatial inversion and time reversal symmetry, it is necessary to break one or both of these symmetries to obtain WPs.

\begin{figure*}[]
    \centering
    \includegraphics[width=\linewidth]{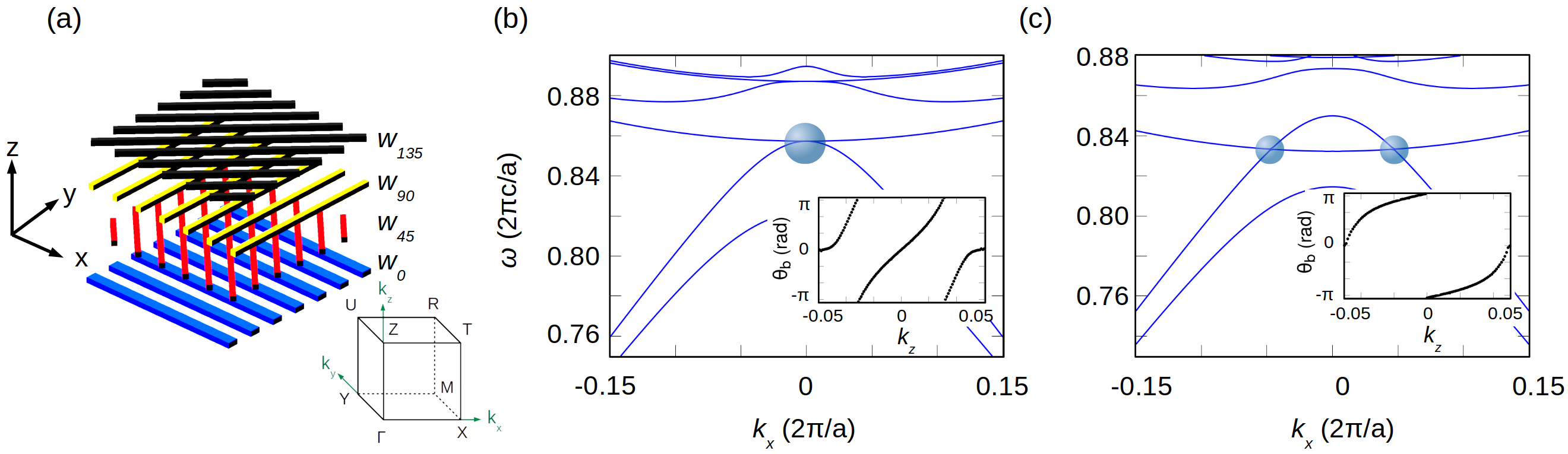}
    \caption{ \small{
        (a)  An exploded view of the chiral woodpile PhC (one unit cell in $z$) and its Brillouin zone. (b) Band structure of the chiral woodpile PhC, made out of dielectric rods ($\varepsilon = 2.31$), with equal widths for all rods at $k_y = k_z = 0$ showing bands 3 to 8. The quadratic WP at $\mathbf{\Gamma}$ is marked with a blue circle. The inset shows a plot of the Berry phase for band 4 around $\mathbf{\Gamma}$. The double winding indicates that the degeneracy has a charge of 2. (c) The band structure of the same chiral woodpile PhC in (b) but with increased rod width $w_0$ in the blue layer of each unit cell. The linear WPs that occur due to the splitting of the quadratic WP in (b) are marked with blue circles. The inset shows a plot of the Berry phase for band 4 around one of the degeneracies. The single winding indicates that the degeneracy has a charge of 1.
      \label{fig:1}}}
\end{figure*}

\begin{figure}[]
    \centering
    \includegraphics[width=\linewidth]{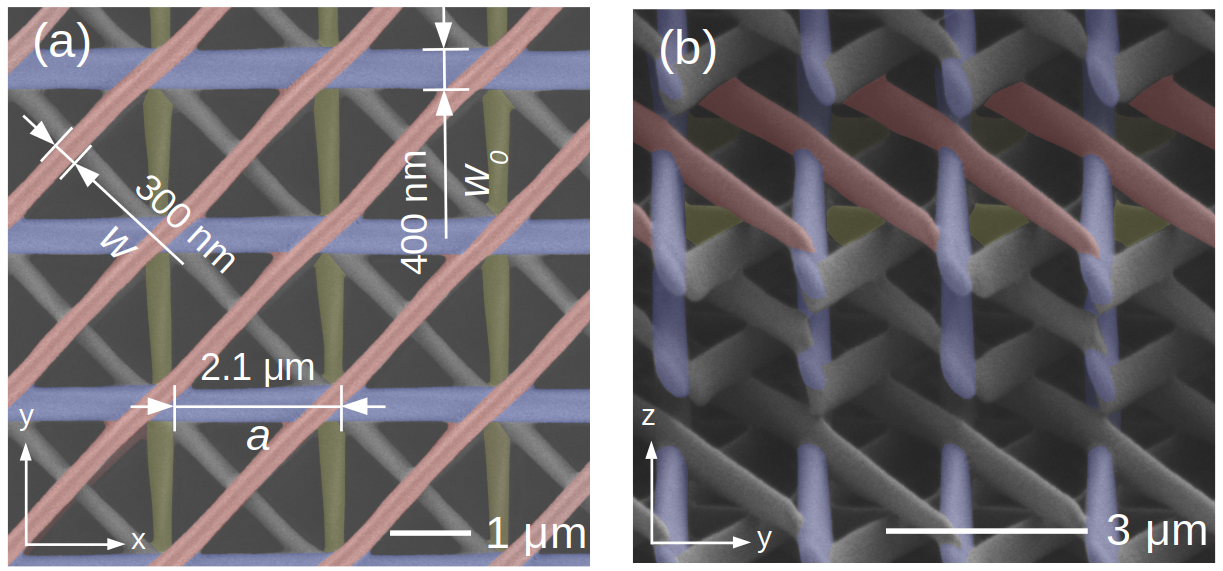}
    \caption{ \small{
        False color SEM image of a typical chiral woodpile photonic crystal with increased rod width $w_0$ in the blue rods. (a) Top view, (b) side view.
      \label{fig:2}}}
\end{figure}

A Weyl point can possess an arbitrary integer charge, $m$, and can be described by the generic three-dimensional Hamiltonian of the form \cite{PhysRevLett.108.266802},
\begin{align}
    \mathcal{H}(\mathbf{k}) = k_+^{|m|}\sigma_+ + k_-^{|m|}\sigma_- + k_z\sigma_z + \omega_0 I
\end{align}
where $k_{\pm} = (k_x \pm ik_y)$, $\sigma_{\pm} = 1/2(\sigma_x \pm i\sigma_y)$, $\sigma_{x,y,z}$ are the Pauli matrices, $I$ is the identity matrix, and $\omega_0$ is the frequency at which the WP occurs in the spectrum. The eigenvalues of $\mathcal{H}(\mathbf{k})$ describe the two bands in the vicinity of the WP and are given by
\begin{align}
    \lambda_{\pm} = \omega_0 \pm \sqrt{(k_x^2+k_y^2)^{|m|} + k_z^2}.
    \label{equ:lambda_WP}
\end{align}
From equation~(\ref{equ:lambda_WP}) it is evident that the leading order $k_{x,y}$-dependence in the dispersion around a WP is governed by its charge and henceforth, charge-1 WPs will be referred to as ``linear" and charge-2 WPs as ``quadratic".

While charge-1 WPs have been extensively studied, higher-charged WPs have received relatively less attention. These exotic WPs, sometimes referred to as unconventional WPs, have markedly different properties than their conventional counterparts such as higher-order algebraic dispersion, multiple Fermi-arc surface states and large density of states for high charge ($m \geq 2$). Higher-charged WPs are known to occur in non-symmorphic crystals due to the stabilization of multiple linear WPs at high symmetry points in momentum space \cite{PhysRevB.95.125136, PhysRevMaterials.5.054202, PhysRevB.98.214110}. In photonics, quadratic WPs were predicted \cite{JPSJ.87.123401} and observed in all-dielectric woodpile-like structures \cite{Sachin}. Such structures exhibit several additional symmetries and therefore provide an ideal platform for tuning the locations of linear WPs that can form from splitting a quadratic WP due to symmetry breaking~\cite{PhysRevB.98.214110,PhysRevB.95.201102}. This tunability of WPs could be of importance in the realization of three-dimensional large-volume, single-mode lasing devices that rely on the vanishing photonic density of states at frequency-isolated linear WPs \cite{Chua:14}. 

In this article, we experimentally demonstrate that under careful symmetry breaking, quadratic WPs can be split into two linear WPs of the same charge. We show by analyzing the underlying symmetries of the structures that for certain choices of defects this splitting occurs strictly along high symmetry directions in momentum space. We also show that the momentum-space separation of the resulting linear WPs can be readily controlled via geometric parameters of the structure. While WPs have mostly been observed in large-scale structures with mm- or cm-scale lattice constants \cite{Yang1013,Lu622,li_weyl_2018}, our platform of choice is a micron-scale three-dimensional photonic crystal (PhC) with low dielectric contrast which is fabricated by a two-photon polymerization process \cite{deubel_direct_2004}. This allows for the realization of WPs at near-infrared wavelengths that are characterized using Fourier transform infrared (FTIR) spectroscopy.

\begin{figure*}[t]
    \centering
    \includegraphics[width=\linewidth]{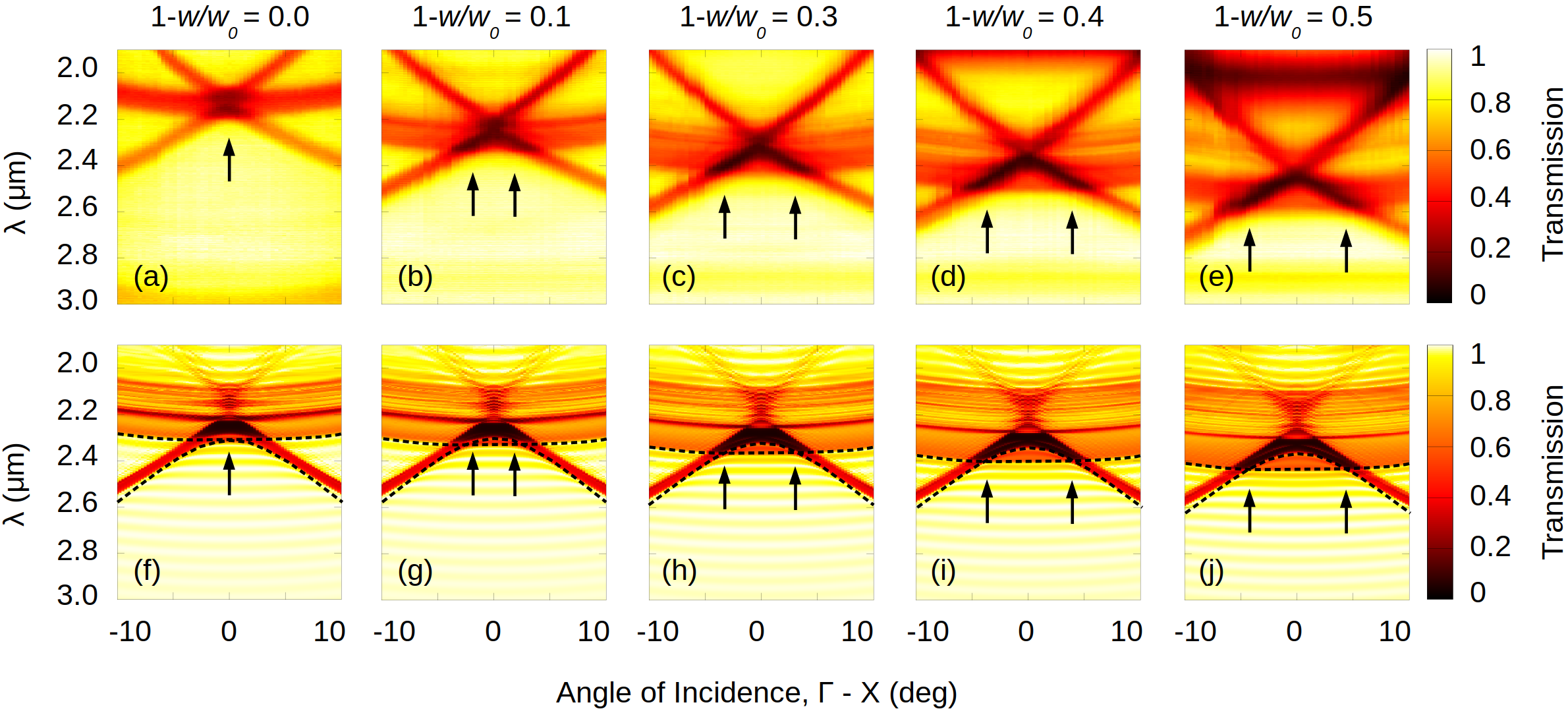}
    \caption{ \small{ (a)-(e) Measured angle-resolved FTIR spectra of the chiral woodpile PhC with varying values of the width $w_0$. The locations of the WPs are marked with arrows. (f)-(j) The corresponding RCWA simulated spectra of the chiral woodpile PhC. The dashed black lines are the $k_z = 0$ bulk bands calculated from MPB.
      \label{fig:3}}}
\end{figure*}

\section{Results}
The particular structure that we employ to realize this phenomenon is a chiral woodpile PhC whose unit cell consists of stacked and partially overlapping layers of rods that have a relative $45^{\circ}$ in-plane rotation between them as shown in Fig.~\ref{fig:1} (a). The lattice constant in all three directions is $a=\SI{2.1}{\um}$. The four rods in the unit cell have height $h$ and widths $w_{0}$, $w_{45}$, $w_{90}$, $w_{135}$ where the subscripts identify the rods by their angle of orientation with respect to the $x$-axis. The rods are made out of a nearly lossless, non-magnetic dielectric material of dielectric constant $\varepsilon = 2.31$ (see Methods). Due to the chirality of this structure, inversion symmetry is broken which allows WPs to exist. Furthermore, when all rod widths and heights are equal, this structure belongs to the non-symmorphic space group $P4_222$ (\# 93), which has a screw axis along the $z$ direction. This screw symmetry, defined by a $90^{\circ}$ rotation in the $x$-$y$ plane and $a/2$ shift along the $z$ direction, results in the presence of a quadratic WP at the Brillouin zone center ($\mathbf{\Gamma}$), as shown in Fig.~\ref{fig:1}, (b) and corners ($\mathbf{R}$).

Instead, when the rod width in one layer in the unit cell is taken to be different from the widths of the other three layers, the screw symmetry is broken, resulting in a splitting of the quadratic WP into two linear WPs as shown in Fig.~\ref{fig:1} (c). For example, when a width defect is introduced such that $w_{45}= w_{90}= w_{135}=w$ and $w_0 \ne w$, the space group of the structure is reduced to $P222$ (\# 16). While this subgroup does not contain the screw symmetry, it retains $C_2$ rotations about the $x$, $y$ and $z$ axes that restrict the splitting directions of the WPs to high symmetry lines in momentum space. This can be argued as follows: Let the splitting associated with this change in space group by a small perturbation result in two linear WPs, one of which is located at a generic momentum point $(k_x,\, k_y,\, k_z)$. The location of both WPs can be inferred from the symmetry operations of the space group $P222$ and time reversal symmetry as they must either map to themselves or to each other under these operations. Time reversal symmetry requires a WP to be located at $(-k_x,\, -k_y, \, -k_z)$. However, the aforementioned $C_2$ rotations require a WP to be located at $ (-k_x,\, k_y, \, -k_z)$, $(k_x,\, -k_y, \, -k_z)$ and $(-k_x,\, -k_y,\, k_z)$. Since there are only two linear WPs that can occur from the splitting of a single quadratic WP, at least two components of these momentum vectors must be set to $0$ or $\pi/a$ such that they map to themselves under a sign flip. As a result, for the quadratic WP at $\mathbf{\Gamma}$, the symmetry-consistent directions of splitting are $\mathbf{\Gamma - X}$, $\mathbf{\Gamma - Y}$ or $\mathbf{\Gamma - Z}$, and for the quadratic WP at $\mathbf{R}$, the symmetry-consistent directions of splitting are $\mathbf{R - U}$, $\mathbf{R - T}$ or $\mathbf{R - M}$. A similar analysis shows that the splitting can be made to occur along diagonal directions in momentum space (e.g. along $\mathbf{\Gamma - M}$ for the WP at $\mathbf{\Gamma}$) when $w_{45}$ or $w_{135}$ is chosen to be different from the other three rod widths. In the most general case where all four rod widths are different, the space group is further reduced to $P2$ (\#3) which retains $C_2$ rotation about the $z$ axis. This still restricts the splitting to occur in the $k_z = 0$ plane or along $\mathbf{\Gamma - Z}$ for the quadratic WP at $\mathbf{\Gamma}$ and $k_z = \pi/a$ plane or along $\mathbf{R - M}$ for the quadratic WP at $\mathbf{R}$. For the remainder of this article, we will focus on the case where a width defect is introduced in a single layer in each unit cell such that $w_{45}= w_{90}= w_{135}=w$ and $w_0 \ne w$.

\begin{figure}[]
    \centering
    \includegraphics[width=\linewidth]{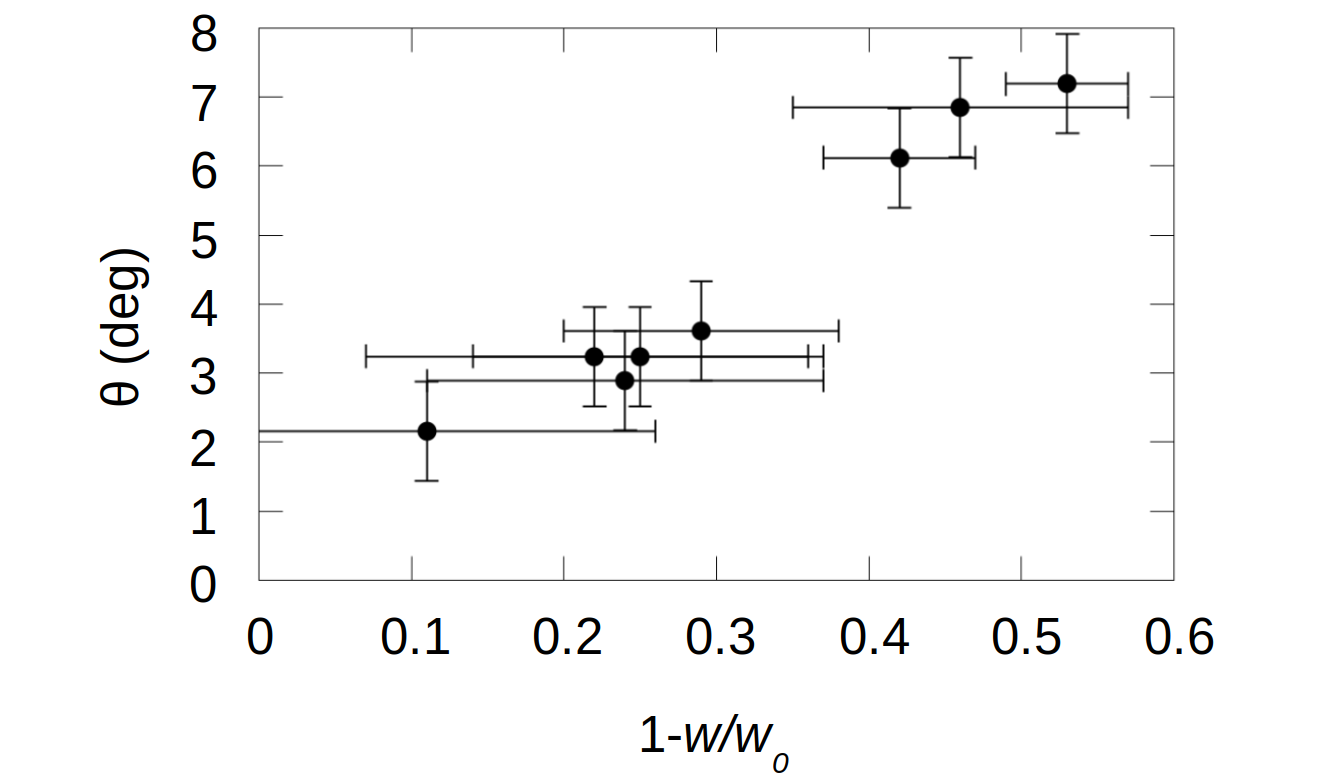}
    \caption{ \small{
        The angular separation of WPs, $\theta$, as a function of the symmetry-breaking parameter, $1-w/w_0$. All data points correspond to samples that were fabricated with the same value of $w$.
      \label{fig:4}}}
\end{figure}

To demonstrate the topological nature of WPs in the split and unsplit cases, we directly calculate their topological charges from the electromagnetic eigenmodes extracted from \textsc{MIT Photonic Bands} (MPB) \cite{Johnson:01}. Since WPs act as sources or sinks of Berry curvature, their charge can be obtained by integrating the Berry curvature over a closed sphere enclosing the WP in momentum space. Alternatively, this can also be calculated using a series of line integrals of Berry connection on closed contours that are sampled from such a sphere. These line integrals represent the geometric phase, called the Berry phase ($\theta_B$), acquired by the eigenmodes as they are adiabatically transported around the closed contours. In the discrete form, $\theta_B$ on one such contour ($C$) is given by \cite{Resta_2000} 
\begin{equation}
    \theta_B(C) = - \textrm{Im} \ln\left(\prod_{i = 1}^{N}\frac{\braket{\psi(\mathbf{k}_{i})|\psi(\mathbf{k}_{i+1})}}{\left|\braket{\psi(\mathbf{k}_{i})|\psi(\mathbf{k}_{i+1})}\right|} \right)
\end{equation}
where the index $i$ discretizes the contour $C$ into $N$ points and $\ket{\psi(\mathbf{k}_i)}$ is the periodic part of the magnetic field eigenmode of the PhC at the momentum point $\mathbf{k}_i$. These contours can be chosen to lie parallel to the $k_x$-$k_y$ plane and $\theta_B$ can therefore  be plotted as a function of $k_z$. The topological charge of the WP is the winding number of $\theta_B (k_z)$. The plots for $\theta_B$ for the split and unsplit cases are shown in the insets of Fig.~\ref{fig:1} (b)-(c) which confirm that the degeneracy at $\mathbf{\Gamma}$ is a quadratic WP of charge $+2$ and the split degeneracies along the $\mathbf{\Gamma-X}$ direction are linear WPs of charge $+1$ each.

For the experiment, we fabricate chiral woodpile PhCs via two-photon lithography of a liquid negative-tone photoresist (IP-Dip, refractive index = 1.52) using a Nanoscribe professional GT. We print several samples of this PhC with a lattice constant of \SI{2.1}{\um} with varying values of $w_0$, while fixing $w$ to \SI{0.2}{\um} for the symmetry broken samples. Due to the voxel's height in $z$, adjacent layers overlap by approximately 50\% of the rod height. A scanning electron microscope (SEM) image of a typical sample is shown in  Fig.~\ref{fig:2}. For characterizing the fabricated PhC, we measure the angle-resolved transmittance via Fourier-transform infrared spectroscopy (FTIR). Details about the fabrication and measurement setup can be found in the Methods section.

For the case of all equal rod widths, we observe a quadratic WP at \SI{2.2}{\um} wavelength at the $\mathbf{\Gamma}$ point (Fig.~\ref{fig:3} (a)). As previously described, an increase in $w_0$ splits the WP along the $\mathbf{\Gamma}-\mathbf{X}$ direction into two linear WPs. This splitting can be seen as a spectral feature corresponding to the two involved bands piercing through each other with the linear WPs occurring at their intersection points as shown in Fig.~\ref{fig:3} (b)-(e). Furthermore, the angular separation between the two linear WPs increases with $w_0$. Along the orthogonal direction, $\mathbf{\Gamma}-\mathbf{Y}$, the separation of the same spectral feature indicates that the two bands move apart and are non-degenerate (see Supplementary Material). To compare our experimental results with theory, we also perform RCWA (rigorous coupled-wave analysis) simulations as implemented in \textsc{Stanford Stratified Structure Solver} (S$^4$) \cite{S4}, to obtain the transmission spectrum of this PhC. The simulated spectra are plotted in Fig.~\ref{fig:3} (f)-(j) and show an excellent agreement with the experimentally obtained data. Moreover, the sharp spectral features match the $k_z = 0$ Weyl bands obtained from MPB (dashed lines in Fig.~\ref{fig:3} (f)-(j)) allowing for a direct observation of the splitting. 

We extract the angular separation $\theta$ between the two linear WPs from the experimental data and plot it as a function of the symmetry-breaking parameter $1-w/w_0$ in Fig.~\ref{fig:4}, demonstrating the tunable nature of WP splitting in this structure. In our experiment, the separation of WPs in momentum space is found to be a monotonic function of the symmetry-breaking parameter, however we note that this is not true in general. For higher dielectric contrast, the separation of the WPs can first increase and then decrease to zero for increasing values of the symmetry-breaking parameter. This can lead to the formation of accidental quadratic WPs which are not symmetry protected but are obtained on the fine tuning of parameters but are otherwise identical to the symmetry-protected quadratic WPs (see Supplementary Material for an example).

At low dielectric contrast, the WPs are embedded inside the $k_z$-projected bands of the PhC which could obscure the dispersion features of the WPs in the spectrum. However, we observe that the $k_z = 0$ Weyl bulk bands are clearly visible as boundaries of sharp features in the spectrum as indicated by the dashed lines in Fig.~\ref{fig:3} (f)-(j). To explain this observation, we perform a coupling analysis similar to \cite{Sachin, CD1,CD2,CD3,CD4} which reveals that the relevant projected bands are nearly of either s- or p- character and hence couple selectively to light of these polarizations (see Supplementary Material). This selective coupling results in the boundary of the projected bands, corresponding to $k_z = 0$ bulk bands, appearing as sharp drops in the transmission of polarized light allowing for the observation of WPs despite the lack of a local bandgap.


Perhaps the most direct physical manifestation of the non-trivial topology of a WP is the presence of surface states that form Fermi arcs connecting WPs of opposite charges. Interestingly, this property is closely related to that of unidirectional edge states found in two-dimensional Chern insulators. We now present a numerical exploration of the surface states associated with both the quadratic and linear WPs in our structure. Since the WPs of interest occur near the Brillouin zone center, the surface states lie above the light line of air. As a result, they are leaky resonances that radiate away from the surface with a finite lifetime. At low dielectric contrast such as in our experiment, the lifetime of such resonances can be so low that they are effectively unobservable in experiment and hard to extract and analyze numerically. To overcome this difficulty, we consider an interface between two chiral woodpile PhCs with opposite handedness for our analysis. This leads to a doubling of the number of surface states due to the presence of WPs on both sides of the interface. Moreover, the surface states originating from both structures are degenerate in frequency and momentum and can therefore hybridize to form states with even and odd symmetry with respect to the mirror plane at the interface. Nevertheless, the topological origin of these surface states can be directly confirmed by examining the surface spectrum along a circular contour that encloses the projection of the WPs. 

\begin{figure}[]
    \centering
    \includegraphics[width=\linewidth]{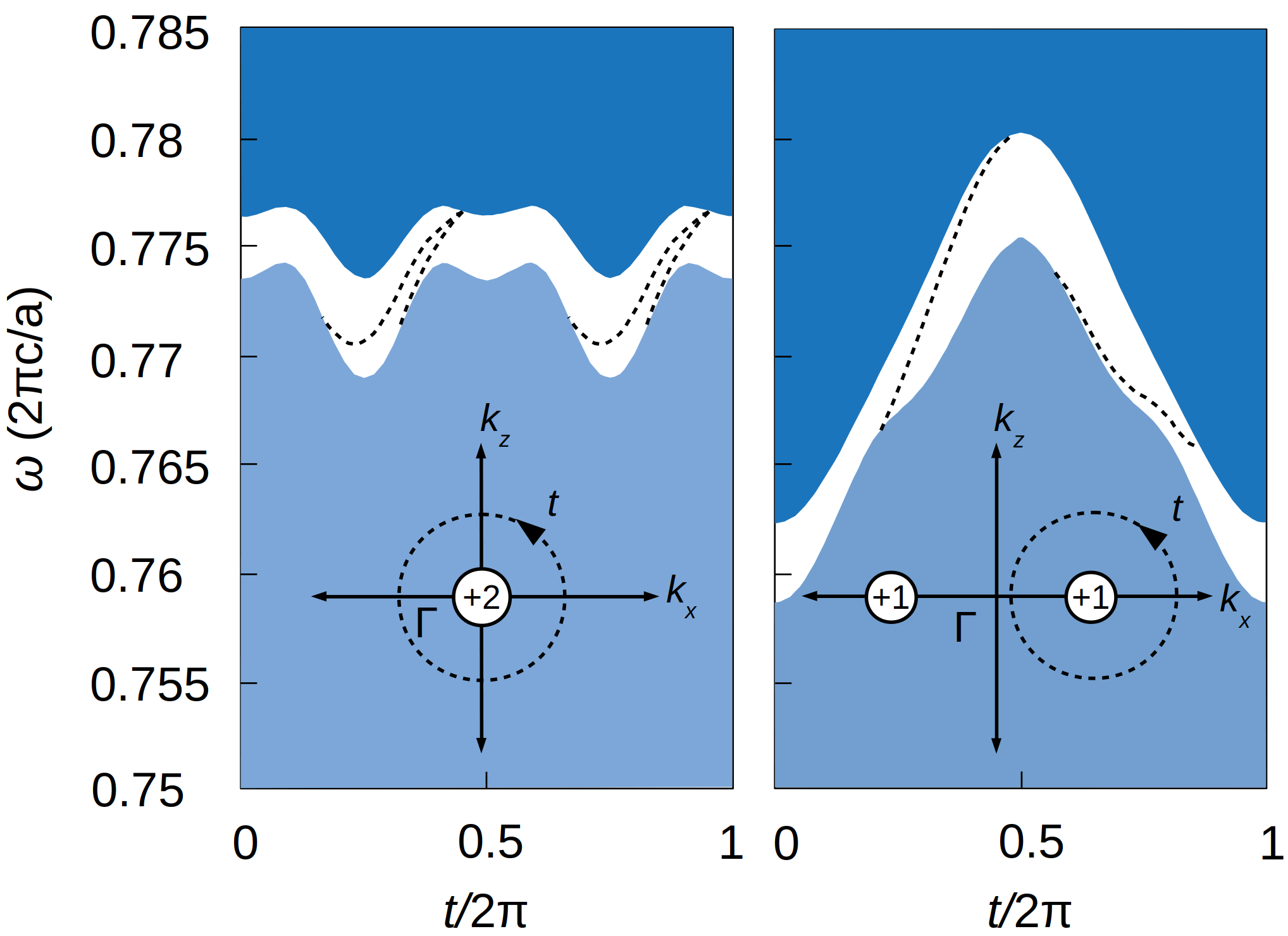}
    \caption{
        Surface states associated with the WPs at the interface between two low-contrast chiral woodpile PhCs with opposite chirality. (a) The $k_y$-projected band structure along the circular loop, parameterized by $t$, enclosing the quadratic WP at $\mathbf{\Gamma}$. The non-trivial gap formed along the loop contains two pairs of hybridized surface states (dashed lines) with even and odd symmetry with respect to the mirror plane at the interface. The blue solid colors are projections of the bulk bands. (b) The $k_y$-projected band structure along the circular loop enclosing one of the split linear WPs, showing only a single pair of even and odd surface states. 
      \label{fig:5}}
\end{figure}

We consider a system consisting of two chiral woodpile PhCs with opposite chirality and equal rod widths in all layers that meet at an interface parallel to the $y = 0$ plane. The surface band structure calculated along a circular loop in the $k_x$-$k_z$ plane enclosing the quadratic WP at $\mathbf{\Gamma}$ is shown in Fig.~\ref{fig:5} (a). This loop has a radius of 0.05 $(2\pi/a)$ and is parametrized by a single angular variable $0 \le t/(2\pi) < 1$. The surface band structure reveals the existence of two pairs of hybridized states that cross the non-trivial gap formed along the loop. Due to the bulk-boundary correspondence principle, the number of surface states implies that the magnitude of charge of the enclosed WPs is 2. Next, we consider the same simulation but for an increased rod width $w_0$ that splits the quadratic WP. The surface band structure along a loop of radius 0.033 $(2\pi/a)$ enclosing just one of the linear WPs is shown in Fig.~\ref{fig:5} (b) which reveals only one pair of surface states in the gap formed along the loop. This indicates that the charge of the enclosed WP has magnitude of 1.

In conclusion, we have observed the splitting of a quadratic Weyl point into two linear Weyl points in a low-contrast 3D PhC. We find that the splitting can be made to occur strictly along high-symmetry directions in momentum space, a consequence of controlled symmetry breaking and that their separation can be readily tuned via the
geometric parameters of the crystal. The micron-scale periodicity of our structure allows us to access Weyl points in the near-infrared optical spectrum. Our approach opens new avenues for designing large-volume single-mode lasers \cite{Chua:14}, using Weyl points and Fermi arc surface states in microscale photonic structures, relevant to near-infrared optics.	

\section{acknowledgments}
C.J. gratefully acknowledges funding from the Alexander von Humboldt Foundation within the Feodor-Lynen Fellowship program. G.v.F. acknowledges funding by the Deutsche Forschungsgemeinschaft through CRC/Transregio 185 OSCAR  (project No.\ 277625399). We would like thank the Nano Structuring Center Kaiserslautern for technical support. M.C.R., S.V.\ and A.C.\ acknowledge the support of the U.S.\ Office of Naval Research (ONR) Multidisciplinary University Research Initiative (MURI) under Grant No.\ N00014-20-1-2325. A.C. acknowledges support from the Center for Integrated Nanotechnologies, an Office of Science User Facility operated for the U.S. Department of Energy (DOE) Office of Science, and the Laboratory Directed Research and Development program at Sandia National Laboratories. Sandia National Laboratories is a multimission laboratory managed and operated by National Technology \& Engineering Solutions of Sandia, LLC, a wholly owned subsidiary of Honeywell International, Inc., for the U.S. DOE’s National Nuclear Security Administration under contract DE-NA-0003525.  The views expressed in the article do not necessarily represent the views of the U.S. DOE or the United States Government.

\section{Methods}
\subsection*{Fabrication}
For the fabrication of the chiral photonic woodpile samples in the IP-Dip resist we use a Nanoscribe Professional GT at a scan speed of \SI{20}{mm/s} and laser power of 62\%, which corresponds to \SI{34}{mW} on the entrance lens of the objective. The structures are printed onto Menzel cover slips (borosilicate glass). Since we use the dip-in configuration of the Nanoscribe the cover slips need to be coated with approximately \SI{13}{nm} of $\mathrm{Al}_2\mathrm{O}_3$ in order to facilitate interface finding. The coated cover slips show a transmission of greater than 75\% for all wavelengths used in our measurements. After printing, the sample is developed for \SI{10}{min} in PGMEA and \SI{10}{min} in isopropanol, subsequently. It is then transferred to a solution of \SI{150}{mg} Irgacure 651 in \SI{24}{ml} of isopropanol and  illuminated for 60s with UV light from an Omnicure S2000 with 95\% iris opening. This post-print UV curing~\cite{Oakdale:16} increases the stability of the woodpile structure. In the end, the sample is blow-dried in a stream of nitrogen.
The complete footprint of the structure is approximately \SI{1}{mm^2} with 20 layers in height. To achieve such a large footprint within reasonable writing time we use elaborate stitching: The structure is printed in portions of 4x4 angled blocks between which the stage is moved for larger travel distance. Inside each block and layer we print the sample using the galvanometer-scanning in combination with piezo-stitching for reduced vignetting and more precise positioning. The alignment of the stage, piezo and galvo axes is ensured by employing the axis transformation implemented in NanoWrite.
The structural parameters are determined via scanning electron microscopy and are listed in the Supplementary Material. While the usual rods consist of just one printed line, the defect rod width is increased by printing multiple lines at a hatching distance between \SI{10}{nm} and \SI{50}{nm}. For the structure in Fig.~\ref{fig:3}b) the laser power used for printing the $w_0$- rods was decreased to 55\%.
\subsection*{Measurement} \label{measurement}
For the measurement of the spectra of the woodpile samples we use the Hyperion 3000 microscope attached to a Bruker Vertex v70 FTIR. The spectra are taken in transmission mode with a halogen lamp and a MCT detector cooled by liquid nitrogen. To increase $k$-space resolution the lower 15x Cassegrain objective is covered except for a pinhole of \SI{1}{mm} in diameter, such that we obtain a nearly collimated beam. We estimate the spread of this beam to be approximately $\pm 0.3$° in the direction along which we measure (the $k_x$ direction in Fig.~\ref{fig:3}), and approximately $\pm 2$° perpendicular to that \cite{cerjan2021observation}.
The sample is then tilted with respect to the beam along its $x$-axis, around perpendicular incidence of the beam. This is done by tilting the sample holder in steps of approximately 0.4°. As we cannot determine the position of perfect perpendicular incidence from the sample positioning (within an error of approximately 5°), we take spectra for both positive and negative tilting angles, and determine $\Gamma$ from the symmetry of the measured angle resolved transmittance spectra.
All spectra are referenced to the transmission of the used substrates, and individually scaled to their maximum at each angle. For each angle we average over 64 measurements with an FTIR resolution set to 4/cm in wavelength. The small dip in transmittance around \SI{2.8}{\um} wavelength, constant across angles, is due to the absorption in the IP-Dip \cite{Fullager:17}. The spectra are post-processed to remove fringes that appear in the spectra due to multiple reflections in the glass and $\mathrm{Al}_2\mathrm{O}_3$.

\bibliography{main_charge2_WPs}

\section*{Author contributions}
C.J. and S.V. contributed equally to this work. S.V. and J.N. conceived the main idea of the project. S.V. and J.N. performed the theoretical analysis and numerical simulations with the assistance of A.C. C.J. fabricated the structures and performed all measurements with help from S.A. M.C.R. and G.v.F. supervised the project. All authors discussed the results and contributed significantly to the preparation of the manuscript.

\section*{Competing interests}
The authors declare no competing interests.
\section*{Additional information}
Correspondence and requests for materials should be addressed to C.J., S.V., or M.C.R.

\end{document}


\title{Supplemental Material for: Observation of Quadratic (Charge-2) Weyl Point Splitting\\ in Near-Infrared Photonic Crystals}
\author{Christina J\"org$^{1,2,\dag,*}$, Sachin Vaidya$^{2,\ddag,*}$, Jiho Noh$^{2,3}$, Alexander Cerjan$^{2,4,5}$, Shyam Augustine$^{1}$, Georg von Freymann$^{1,6}$, Mikael C. Rechtsman$^{2}$}
\affiliation{
 $^1$Physics Department and Research Center OPTIMAS, Technische Universität
Kaiserslautern, 67663 Kaiserslautern, Germany\\
 $^2$Department of Physics, The Pennsylvania State University, University Park, PA 16802, USA\\
 $^3$Department of Mechanical Science and Engineering, University of Illinois at Urbana-Champaign, Urbana, Illinois 61801, USA\\
 $^4$Sandia National Laboratories, Albuquerque, New Mexico 87123,  USA\\
 $^5$Center for Integrated Nanotechnologies, Sandia National Laboratories, Albuquerque 87123, New Mexico, USA\\
 $^6$Fraunhofer Institute for Industrial Mathematics ITWM, 67663
Kaiserslautern, Germany\\
 $^*$These authors contributed equally to this work.
}

\def\thefootnote{}\footnotetext{$^\dag$ christinaijoerg@gmail.com, $^\ddag$ sxv221@psu.edu}

\date{\today}

\maketitle

\setcounter{equation}{0}
\renewcommand{\theequation}{S{\arabic{equation}}}
\setcounter{figure}{0}
\renewcommand{\thefigure}{S{\arabic{figure}}}

\tableofcontents

\newpage

\section{Measurements and simulations along $\mathbf{\Gamma-Y}$\label{sec:GY}}
\begin{figure*}[h!]
    \centering
    \includegraphics[width=\linewidth]{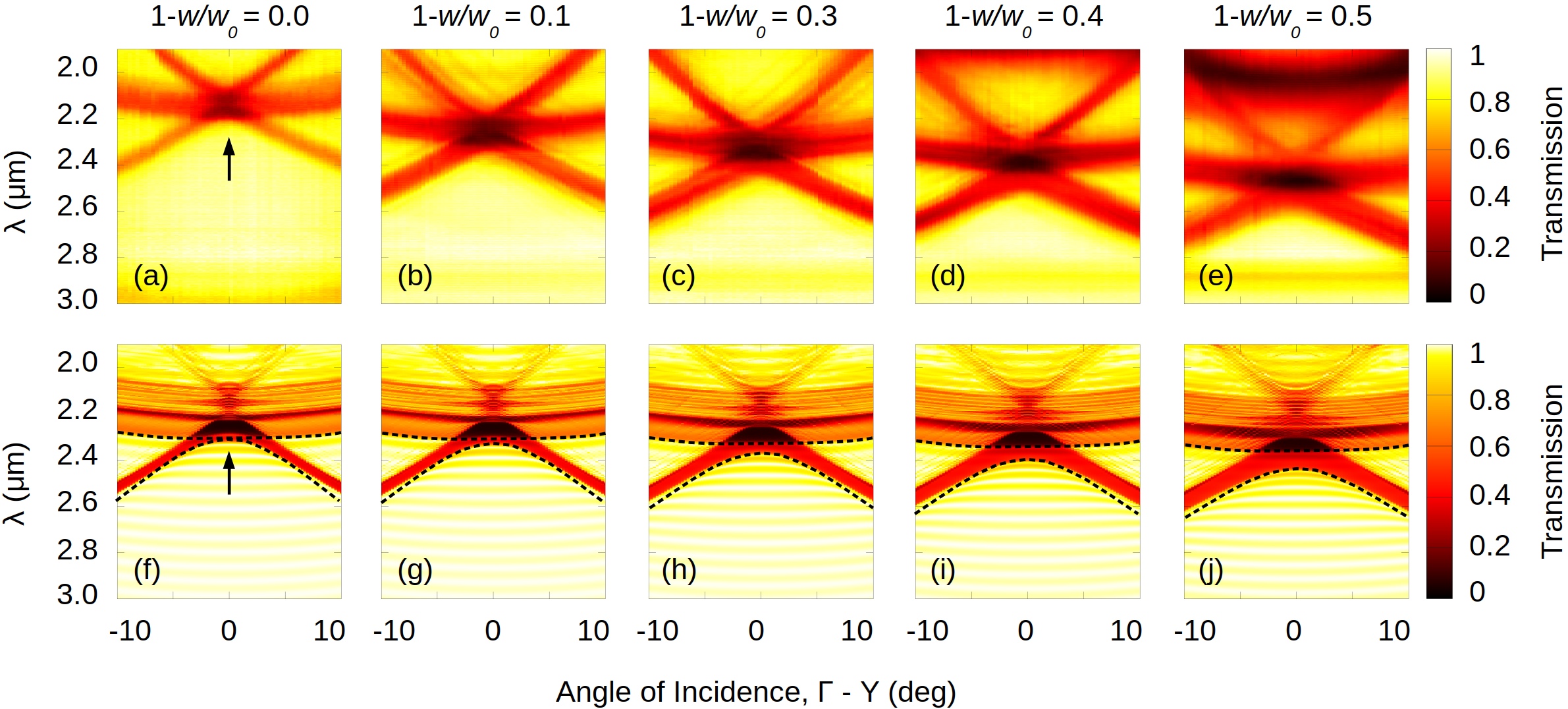}
    \caption{ \small{
Spectra along $\mathbf{\Gamma - Y}$: 
 (a)-(e) Measured angle-resolved FTIR spectra of the chiral woodpile PhC with varying values of the width $w_0$. (f)-(j) The corresponding RCWA simulated spectra of the chiral woodpile PhC. The dashed black lines are bulk bands calculated from MPB. We see that the two bands that form the WP for $w=w_0$ move apart for increasing rod width $w_0$.
      \label{fig:S1}}}
\end{figure*}

In addition to the spectra taken along the $\mathbf{\Gamma - X}$ direction, we also performed measurements and simulations along $\mathbf{\Gamma - Y}$, which are shown in Fig.~\ref{fig:S1}. We see here that the two bands, which form the quadratic WP for equal rod widths in all layers, move apart upon increasing $w_0$ and are no longer degenerate. This is as expected since the linear WPs are point degeneracies that only occur along the $\mathbf{\Gamma-X}$ direction.

\section{Sample parameters \label{sec:sampleParam}}

The structural parameters of the woodpile PhC samples, as determined by SEM, are given in table~\ref{table:1}
.
\begin{table}[h!]
\centering
\begin{tabular}{|c|c|c|c|} 
 \hline
 Fig. & $w$ / µm & $w_0$ / µm & $a$ / µm \\
 \hline\hline
 3a)+4 & \SI{0.15\pm 0.01}{} & \SI{0.14 \pm 0.01}{} & \SI{2.0 \pm 0.1}{} \\ 
 3b)+4 & \SI{0.20\pm 0.02}{} & \SI{0.22 \pm 0.01}{} & \SI{2.1 \pm 0.1}{} \\ 
 3c)+4 & \SI{0.19\pm 0.01}{} & \SI{0.26 \pm 0.02}{} & \SI{2.1 \pm 0.1}{} \\ 
 3d)+4 & \SI{0.20\pm 0.01}{} & \SI{0.35 \pm 0.01}{} & \SI{2.1 \pm 0.2}{} \\ 
 3e) & \SI{0.23\pm 0.02}{} & \SI{0.43 \pm 0.05}{} & \SI{2.0 \pm 0.1}{} \\ 
  \hline
 4) & \SI{0.19\pm 0.03}{} & \SI{0.24 \pm 0.01}{} & \SI{2.1 \pm 0.1}{} \\ 
 4) & \SI{0.19\pm 0.02}{} & \SI{0.25 \pm 0.02}{} & \SI{2.0 \pm 0.1}{} \\ 
 4) & \SI{0.21\pm 0.02}{} & \SI{0.39 \pm 0.02}{} & \SI{2.1 \pm 0.1}{} \\ 
 4) & \SI{0.21\pm 0.01}{} & \SI{0.45 \pm 0.01}{} & \SI{2.1 \pm 0.1}{} \\ 
 \hline
\end{tabular}
\caption{Structural parameters of the samples used in Fig. 3 and Fig. 4 in the main text as determined by SEM.}
\label{table:1}
\end{table}

Horizontal error bars in Fig.~4 of the main text are a result of propagation of the individual errors of $w$ and $w_0$ as given in Table~\ref{table:1}. Vertical error bars in Fig.~4 are assumed to be 2 pixels wide, which corresponds to 0.72°, since we read off the position of the split WPs from the angle-resolved spectra. Therefore, $\theta$ is the difference between the two angles of incidence, as plotted on the horizontal axis in Fig.~3, at which the split WPs are located.

\section{Accidental quadratic Weyl points \label{sec:AccidentalWPs}}
\begin{figure}[h!]
    \centering
    \includegraphics[width=\linewidth]{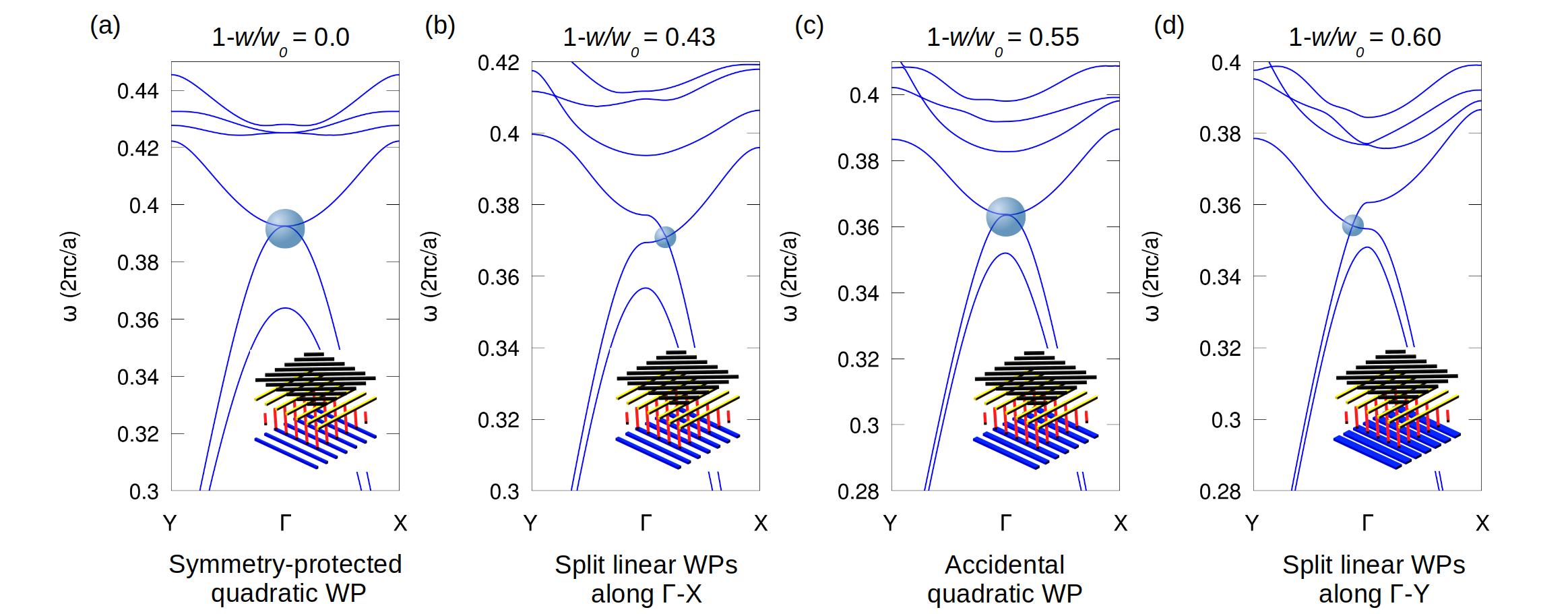}
    \caption{ \small{
(a) Symmetry-protected quadratic WP for $w_0=w$. (b) This quadratic WP first splits into two linear WPs along $\mathbf{\Gamma - X}$ upon increasing $w_0$. (c) Further increasing $w_0$ re-merges the two linear WPs, leading to the formation of an accidental quadratic WP. (d) Increasing $w_0$ beyond the re-merging point leads to a splitting of this quadratic WP along $\mathbf{\Gamma - Y}$.
      \label{fig:S2}}}
\end{figure}

As stated in the main text, the momentum space separation of the WPs is not generally a monotonic function of the symmetry-breaking parameter, $w/w_0$. We present an example of an accidental quadratic WP which is formed due to the re-merging of the split linear WPs on increasing the symmetry-breaking parameter. The band structure for a chiral woodpile made out of Si rods ($\varepsilon = 12$) is shown in Fig.~\ref{fig:S2}. For $w_0=w$ we have the quadratic WP, protected by screw symmetry (Fig.~\ref{fig:S2} (a)). As with the low-contrast PhC in the experiment, upon increasing $w_0$, this quadratic Weyl point splits into two linear WPs, whose separation first increases along the $\mathbf{\Gamma - X}$ direction (Fig.~\ref{fig:S2} (b)). However, increasing $w_0$ further causes their separation to decrease to zero (Fig.~\ref{fig:S2} (c)), leading to the formation of an accidental quadratic WP. Increasing $w_0$ beyond this re-merging point leads to a splitting of this quadratic WP along a different symmetry-allowed direction, $\mathbf{\Gamma - Y}$ in this case (Fig.~\ref{fig:S2} (d)).

\section{Coupling analysis\label{sec:coupling}}
In the experiment presented in the main text, the WPs are embedded inside a continuum of states of other projected bands of the PhC. Despite this, we are able to observe the WPs and $k_z=0$ Weyl bands as the boundaries of spectral features where the transmission drops sharply to zero. Here, we explain this observation by performing a coupling analysis similar to \cite{CD1,CD2,CD3,CD4, Sachin}. Two coefficients, $C_s$ and $C_p$, and mode coupling index, $\kappa$, are defined using the overlap integrals between s-polarized (p-polarized) plane waves $\ket{s(p)}$ and the Bloch modes of the PhC $\ket{\psi_{n,\mathbf{k}}(x,y,z)}$ with momentum $\mathbf{k}$ and band index $n$: 
\begin{equation}
    \eta_{s(p)} = |\braket{s (p)|\psi_{n,\mathbf{k}}(x,y,z)}|^2, \qquad \kappa = \eta_s + \eta_p, \qquad C_{s(p)} = \eta_{s(p)}/\kappa.
\end{equation}

The coefficient $C_{s(p)}$ ($0 \le C_{s(p)} \le 1$) measures the degree of coupling between the polarized plane waves and the Bloch modes of the PhC while $\kappa$ ($0 \le \kappa \le 1$) measures the strength of coupling to a plane wave of any arbitrary polarization. Vanishingly small transmission in the spectrum in the absence of bandgaps can be thought of as either a polarization mismatch between the incident wave and the modes of the PhC, indicated by small $C_{s(p)}$, and/or inefficient mode in-coupling, indicated by small $\kappa$.

\begin{figure}[h]
    \centering
    \includegraphics[width=\linewidth]{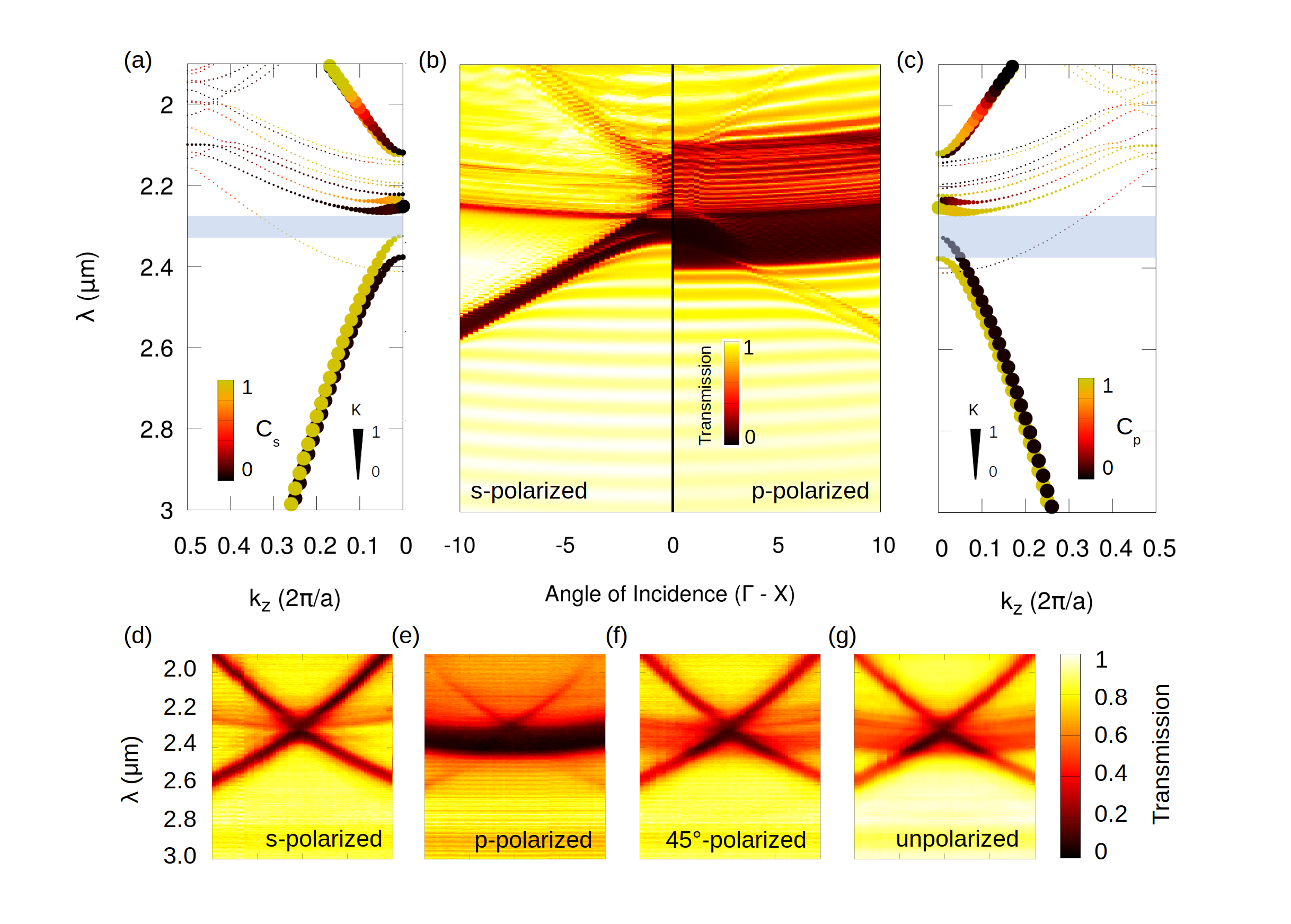}
    \caption{ \small{
(a), (c) The band structure showing bands 3 to 18 for $\mathbf{k} = (0,0,k_z)$. The color of the circular dots indicates the value of $C_{s(p)}$ and their size is proportional to the value of $\kappa$ for the corresponding Bloch modes. The blue highlighted region shows a band of wavelengths with polarization mismatch and/or poor mode in-coupling, leading to the observed sharp drop in transmission at $\mathbf{k_{\parallel}} = \Gamma$. The large-wavelength boundaries of these features correspond to bands 5 and 4 at $k_z = 0$ for s and p polarizations, respectively. (b) The RCWA simulated s- and p-polarized transmission spectra with parameters corresponding to Fig. 3 (h) of the main text. (d)-(g) Experimentally measured transmission spectra for s-polarized, p-polarized, 45$^{\circ}$-polarized and unpolarized light, respectively.
      \label{fig:S3}}}
\end{figure}

In Fig.~\ref{fig:S3}, we analyze the cut of the transmission spectra for each polarization at $\mathbf{\Gamma}$ for a PhC with parameters corresponding to Fig. 3 (h) of the main text. The band structure along the projected momentum $k_z$, for $\mathbf{k_{\parallel}} = \mathbf{\Gamma}$, are shown in Fig.~\ref{fig:S3} (a) and (c). Here, the states are characterized by the values of $C_{s(p)}$ and $\kappa$. The RCWA simulations of the s- and p- polarized spectra are shown in Fig.~\ref{fig:S3} (b). We can see that the sharp drop in transmission at $\mathbf{\Gamma}$ around \SI{2.4}{\um} for both polarizations is a result of a polarization mismatch and/or poor mode in-coupling of the Bloch modes (highlighted in blue in Fig.~\ref{fig:S3} (a) and (c)). The boundaries of these spectral features in each polarization match the two $k_z=0$ Weyl bands, allowing for a direct observation of WPs in the spectrum of $45^{\circ}$-polarized light. In the experiment we measure the spectrum of unpolarized light which is similar to that of $45^{\circ}$-polarized light as can be seen from the experimentally obtained polarized spectrum plotted in Fig.~\ref{fig:S3} (d)-(g). Simulations away from $\mathbf{k_{\parallel}} = \mathbf{\Gamma}$ show that this analysis continues to hold, allowing for this relatively unobscured observation of the splitting of WPs.

\bibliography{si_Weyl_splitting_ref}